# The Cambrian impact hypothesis

Weijia Zhang

Department of Yuanpei Experimental Plan

Peking University

Beijing, PRC 100871

*itaisa@pku.edu.cn*

**Abstract**

After a thorough research on the Earth's circumstantial changes and the great evolution of life in the Cambrian period, the author propounds such a hypothesis: During the Late Precambrian, about 500~600Ma, a celestial body impacted the Earth. The high temperature ended the great glaciation, facilitated the communication of biological information. The rapid change of Earth environment enkindled the genesis-control system, and released the HSP-90 variations. After the impact, benefited from the protection of the new ozone layer and the energy supplement of the aerobic respiration, those survived underground life exploded. They generated carapaces and complex metabolism to adjust to the new circumstance of high temperature and high pressure. That's the Cambrian explosion.

This article uses a large amount of analyses and calculations, and illustrates that this hypothesis fits well with most of the important incidences in astronomic and geologic discoveries.

# 1 Introduction

The Cambrian explosion was the seemingly rapid appearance of most major groups of complex animals around 530 million years ago, as evidenced by the fossil record. This was accompanied by a major diversification of other organisms. It was discovered that the history of life on earth goes back at least 3,550 million years: rocks of that age at Warrawoona in Australia contain fossils of stromatolites, stubby pillars that are formed by colonies of micro-organisms [1]. However, before about 580 million years ago, most organisms were simple, composed of individual cells occasionally organized into colonies. Over the following 70 or 80 million years the rate of evolution accelerated by an order of magnitude.

The seemingly rapid appearance of fossils in the "Primordial Strata" was noted as early as the mid 19th century by Buckland [2], and Charles Darwin saw it as one of the main objections that could be made against his theory of evolution by natural selection [3].

The long-running puzzlement about the appearance of the Cambrian fauna, seemingly abruptly and from nowhere, centers on three key points: whether there really was a mass diversification of complex organisms over a relatively short period of time during the early Cambrian; what might have caused such rapid evolution; and what it would imply about the origin and evolution of animals.

The intense modern interest in this "Cambrian explosion" was sparked by Whittington [4], who in the 1970s re-analysed many fossils from the Burgess Shale (see below) and concluded that several were complex as but different from any living animals.

The explosion was supported by Chenjiang Fauna [5][6]. The Chengjiang Fauna, a well preserved fauna, was found in Chengjiang County of Yun-nan Province in 1984. It is this discovery and subsequent researches that confirm the Cambrian Explosion occurred.

The diversity of many Cambrian assemblages is similar to today's according to Harvey and Butterfield [7, 8].

However, the author did a thorough research on Earth's circumstantial changes and the Cambrian explosion, and was surprised that it's the impact which caused the mysterious changes. Investigations revealed the impact fits well with most of the important incidences in astronomic and geologic discoveries.

# 2 The Cambrian Collisional Life-Evolution and evidences

As our understanding of the events of the Cambrian becomes clearer, data has accumulated to make some hypotheses look improbable. Causes that have been proposed but are now discounted include the evolution of herbivory, vast changes in the speed of tectonic plate movement or of the cyclic changes in the Earth's orbital motion, or the operation of different evolutionary mechanisms from those that are seen in the rest of the Phanerozoic eon.

After a thorough research on the Earth's circumstantial changes and the great evolution of life-forms in the Cambrian Period, to shed light upon a number of issues, the author did propound such a hypothesis:

In the Late Precambrian, about 500~600 million years ago, a celestial body impacted the Earth. The high temperature ended the great glaciation, facilitated the communication of biological information. The rapid changes of Earth's environment enkindled the genesis-control system, and released the HSP-90 variations.

After the impact, benefited from the protection of new ozone layer and the energy supplement of the aerobic respiration, those survived underground life exploded. They generated carapaces and complex metabolism to adjust to the new circumstance of high temperature and high pressure.

## 2.1 The meteor crater and the cycle

It's the most exciting period in Earth history. Before this research, the author has noticed a cycle in Earth history. Maybe that's why the impact occurs.

There are some giant but disputable events. They are called Astroblems——great impact events.

The aftermath of the 4 largest calamities, the considerable and disputable meteorites at 65Ma, 580Ma, 1700Ma and 2300Ma, including atmospheric change, hydrospheric change, biospheric change and lithospheric change, are surprisingly similar.

65 million years ago, the impact event in Late Cretaceous caused the extinction of dinosaurs[9]. In Yucatan, the crater's diameter reached 180 kilometers.

580 million years ago, the Acraman meteor event, whose crater in Australia is confirmed, has a 160km crater diameter. This meteor crater is one of the three largest meteor events ever discovered [10]. The crater is just in the same time with the Cambrian impact.

1.7 billion years ago, Sudbury, the meteor crater's diameter reached 200 kilometers, which is one of the three largest meteor events ever discovered [11].

2.3 billion years ago:"Geological environment (lithosphere, ecosphere, hydrosphere, atmosphere) suffered a catastrophe caused by extraterrestrial diathesis. After that, oxygenic atmosphere formed, life evolution occurred… stromatolites widely developed…carbonatites largely deposited in all continents…" [12].

According to a study group in Peking University [12], it should be a large meteor event.

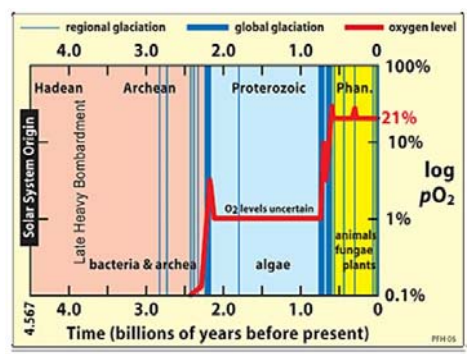

Figure 1: $O_2$'s increase according to Hoffman [13].Life is not the main reason of oxygen's increase. The oxygen's level almost kept constant in Proterozoic and Phanerobiotic. The rapid increase occurred at 2300Ma and 600Ma.

In both confirmed and disputable meteor catastrophes, the four largest ones indicated a cycle of two cosmic years.
65Ma~580Ma~ (Vacancy) ~1700Ma~2300Ma

(We will find the period is decreasing from 600Ma ($23-17=6$) to 535Ma ($6-0.65=5.35$). In fact the cosmic year is decreasing, from 300Ma in 2 billion years ago to the present data of 250Ma [14].)

It is just the time our Solar System needs to pass the four major swing arms once.

We all know the time Solar System needs to move around the galaxy once. It is called a cosmic year. And while the Solar System is moving, the four major swing arms of the galaxy are circling round too. Furthermore, the velocity of the screwy gravitational field was just half of the Solar System's velocity. Hereby the final result is——the time our Solar System needs to pass the four major swing arms once is neither more nor less than 2 cosmic years! [15].

There is an abnormal gravitational origin in one of the major swing arms.

The Solar System is a 2-body system. A planet only has gravitational operation with the Sun. The gravitation between planets is trivial. All 2-body celestial systems have analytic state and its macroscopical state can be calculated. But when the Solar System passes the abnormal gravitational field, its strong force works, resulted in a 3-body celestial system. A 3-body gravitational system is a state of chaos. The state is confused and impacts become possible.

Maybe that's why there are always some "aura events" or "aura stratums" before important catastrophic periods in Earth history. Here is a structure map of the Milky galaxy：

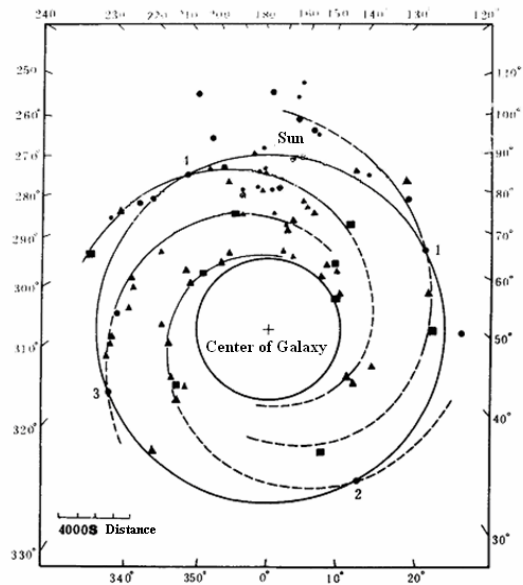

Figure 2: The Sun will meet each major arm at point 1, 2, 3, and 4 in sequence. [15]

## 2.2 High temperature and high pressure caused by impact

In the Late Precambrian, the whole Earth was in a long ice age, and moraines were found in all current continents[16、17、18、19、20]. Only primal one-celled life could live in such a rigorous environment.

First, computer simulations [21、22] pointed out that a giant impact would cause global high temperature and high pressure. In Early Cambrian, the second Neoproterozoic global glaciation ---Marinoan (~600Ma) melted, flood inundated the Earth [23、24], implying a huge heat. That's why there were traces of water scouring on ancient continents.

Second, in the "Cambrian Explosion", metazoan rapidly appeared, formed almost all kinds of life today, and featured by carapaces. The examples suggested high temperature and high pressure, which successfully promoted the Cambrian Explosion [25].

## 2.3 The mechanism of collisional life evolution

The impact destroyed the former and moldering life system. Many advantages created by the giant impact promoted the new life-form's evolution.

There is considerable evidence to support this course.

1) The impact ended the glacier: During the ice age seas were blocked, life couldn't exchange genetic information. Then the collisional high temperature and high pressure ended the glacier, and made the exchange of genetic information possible and convenient.

2) The impact generated oxygen:

Berner and Canfield[26]: Before the Cambrian period the concentration of oxygen in the atmosphere was quite low. (The deposition of Bar Ferrite, the oceanic anoxia, the uranous conglomerate also supported their result.) But in the Early Cambrian, the level jumped to 7%-10% of the current value. Finally, in Late Cambrian, it reached the current level. The ozone increased so rapidly, indicating a tremendous event. Such a sudden increase implied a collisional thermal decomposition of the seawater. After the decomposition, the hydrogen would escape from the atmosphere for its diffusion velocity which is much faster than any other gas. Thus the oxygen was reserved.

Paul G. Falkowski and his research group[27] analyzed the isotopic records of carbon and sulfur in the primal oceanic deposition and got the oxygen level in the past hundreds of million years. The oxygen level at 600Ma was 20%, very close to the current level.

The efficiency of aerobic respiration is 18 times of anaerobic respiration. The energy for an evolution was created.

3) The impact generated ozone layer: The ozone layer appeared in the Cambrian period[28]. Since the impact caused the high temperature and generated enough oxygen, the reactions below would happen: $O_2=2O$, $O_2+O=O_3$. The ozone layer is one of the most effective protections of life.

4) The impact activated the gene-control system: The control gene determined the expression of other genes[29, 30]. And is the key to the Cambrian Explosion.

The earliest control gene ever discovered was in Burgess Shale (535Ma), just in the same time with the impact.

5) The impact activated Heat-shock protein 90 (HSP90): HSP90 accumulates mutations. As soon as the environment suddenly changed, all DNA mutations would express. Then the new life-forms would evolve with different morphosis in a short period. These diversions can be inherited[31].

6) The impact attributed to the development of amphiploids: Because the great change of environment could make the chromosome doubled, the giant impact event would make advantage for amphiploids. Therefore the sexual reproduction was generalized. And it would rapidly promote the life evolution. The Precambrian glacier and the accreted carbonate layer just indicated an acute change of climate[32].

Evidence: 1).Zhang et al.[33]: At 600Ma, the life-form with sexual reproduction appeared. 2).Most of the life in the Cambrian Explosion used sexual reproduction.

7) The impact resulted in a quick evolution of carapaces: 1.With the advantages above, many new kinds of algae developed, contributed to the enrichment of phosphor and prepared enough material for the genesis of carapaces; 2. Since a large amount of ammonia was created (a small part of the hydrogen generated by the thermal decomposition could combine with nitrogen, became the alkaline air, then dissolved in sea and became ammonia), the ocean's pH became alkalescence. Then it's much harder for animals to excrete excrescent resultants out, and carapaces were formed 3.High temperature and high pressure compelled the new life to generate carapaces, or they would be eliminated through selection.

8) The high temperature accelerated the molecular evolution: By C. E. Shannon[34]'s information theory, we got the microcosmic expression of the evolution——an available mutation is a DNA record which reflected the real world more exactly. It works as information. If the amount of the information increases 1bit, the environment should at least input $kT\ln 2$ J entropy into the system. The author defined the entropy an evolution needs as the efficient entropy $S_{efficient}$ and defined the efficiency a species of animal uses entropy from the environment as $\eta$. Therefore,

$$S_{efficient} = S_{environment} \times \eta$$

The collisional high energy brought entropy stream, and η increased 18 times with the aerobic respiration, thus the $S_{efficient}$ was enough for an evolution.

Thus the molecular clock dilemma is solved. The rate of molecular evolution in Early Cambrian is 65 times of current value [35, 36]. The large numerical value can be explained as the result of high temperature and great entropy stream (It accelerated the molecular evolution). The result that the rate of molecular evolution in the Precambrian is much slower than current value [37, 38] can be explained in the same way.

The survived life shall explode freely in a new and favorable world!

## 2.4 The ancient equator and the axis's change driven by stability

In the Late Cambrian, the ancient Australia continent was at the equator, but researchers found the remnant of glacier there [16, 17, 18, 19, 20]. The analyses of moraine also showed the distribution of the glacier included all continents, and were at low latitudes [16, 17, 18, 20, 39, 40, 41].

Accordinly, at that time the axis of rotation is close to the current equator until the impact occurred.

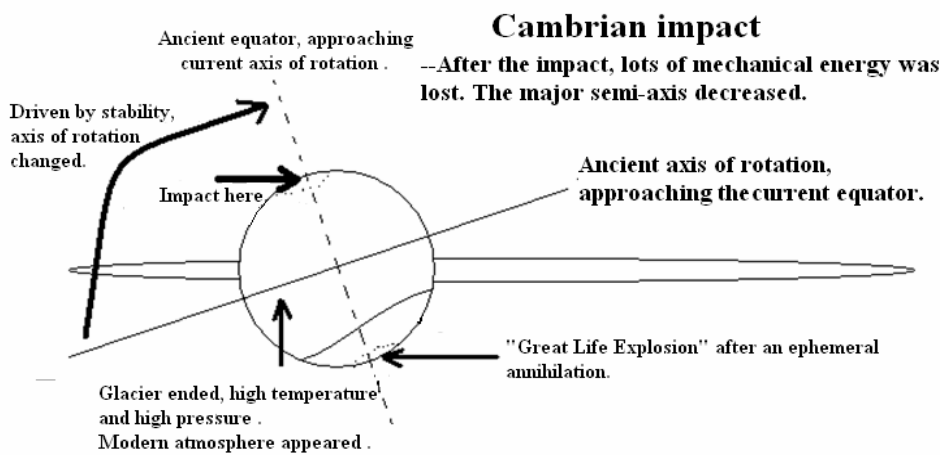

Figure 3: The Cambrian impact suggested and painted by the author.

After a giant impact, the axis of impact would become the axis of rotation, which can be explained by the energy minimum principle of a rotational mode and the parallel axis theorem.（See 3.4.1）

Hays et. al. [42] confirmed the calculation by Milankovitch：the more acclivitous the axis of rotation is, the colder the climate would be.

This can explained the simulation result of Neoproterozoic global glaciation [43].

It is George E. Williams[44] who first put forward the hypothesis that in Late Precambrian the Earth was inclining. Williams disregarded many oppugnations and published his work with courage.

Now, the Cambrian impact hypothesis could explain these oppugnations：

1. The deposition of Bar Ferrite——since the oxygen was suddenly generated by thermal decomposition, the Neoproterozoic atmosphere before the impact is neutral, and the Bar Ferrite could exist.
2. The carbonate cap——the impact would cause high temperature.
3. C13 negative excursion——the impact exterminated the former ecosystem.
4. Once the obliquity of the ecliptic got smaller ,it won't change because of the gravity between the Earth and the Moon（From Hoffman et al. [13]）——the impact just ended this self-stability.
5. Williams's mode cannot explain the high level of ferrite and iridium in the carbonate cap（From Hoffman et al.[13]）——the extraterrestrial impact increased the concentration of ferrite and iridium.

# 3 Further evidences

The hypothesis got evidence which almost supported every process in the impact and can shed light upon most of the unsolved issues in Earth history.

## 3.1. Traces of impact

1). Iridium anomaly

There was iridium anomaly in the Meishucun section where Cambrian explosion was discovered[45, 46]. The Ir concentration measured by China Institute of Atomic Energy in 1985 reached 3.97ppb.

In the middle of Sanxia and Chenjiang, researchers also discovered the iridium anomaly by analyzing the sample of Precambrian-Cambrian boundary [45].

Such anomalies are widely witnessed in China [47].

In many Cambrian layers of China developed black shales with Ni-Mo contained. The iridium anomaly reached 11~13mg/t there. Such shales were also discovered in worldwide regions [48].

2). A Cambrian meteor crater which may be at the same time with the suggested impact

There are three huge meteor craters of which the diameters were ≥160km. Among these, the Acraman meteor crater (580Ma) in Australia is confirmed by Williams [10].

It is the abnormal gravitational field which caused the impact events, primary or concomitant. Maybe the Acraman is a concomitant event.

3). Cambrian impact ejecta deposits in Australia

In the Precambrian shales of Adelaide, south Australia (~600Ma), the impact ejecta deposits can be traced back to 260km [10, 49].

4). Discovery of the iron spherules and silicon particles in the White Clay near Precambrian /Cambrian boundary

In the Meishucun section（P∈-∈）, large number of high-silica contained iron spherules and silicon particles which maybe formed in a high temperature were discovered in the White Clay near Precambrian /Cambrian boundary [46, 47, 50]. They directly indicated an extraterrestrial catastrophic event.

5). Discovery of cosmic dusts

Bodiselitsch [51] found cosmic dusts in the carbonate cap of Marinoan.

## 3.2. Traces of collisional evolution

1).The history of a former life-form, life on the Earth didn't come down in one continuous line

Zhang et al. [52]：in the Precambrian /Cambrian boundary, the concentrations of Dinosterane, C26, C28- sterane and Gammacerane were egregiously high. The critical point is at 540Ma. Before 540Ma, the level is 6~8 times of current value and lasted hundreds of million years; After 540Ma, the concentrations decreased rapidly and reached current level.

The result showed, "This strange distributing indicated that maybe in the Cambrian period or even earlier period lived strange life-forms." [52]

2). Extermination of former life system in the Late Precambrian

The δC13 rapidly decreased in the Early Cambrian, reported by Kirschvink [53], reached the vale, implying an extermination, then rapidly increased, implying the Cambrian Explosion. The analysis of δC13 value's changes in the 8 important boundary lines of Phanerobiotic showed that the range of δC13 negative excursion is in direct proportion with the extent of life extermination.

Study of Meishucun micro-fossils showed that more than 70% families and 80% genera categories exterminated in Early Cambrian [47, 50]. The δC13 negative excursion at the P∈-∈ boundary line in Sanxia section and Meishucun section were discovered by Hsu et al. [45、54].

3). Survivors of the catastrophe and the Cambrian Explosion

Even in such an extraterrestrial catastrophic event, some of the underground animalcule could be protected. They survived and evolved. Karl Stetter and Susan M. Barns[55、56、57、58] have proved that modern life came from those underground ancestors by 16S rRNA tech.

4). Late Cambrian extinction event

When the high temperature dissipated, the dusts still shield the sky. A cold period came. Lots of these nouveaux riches in the warm climate would die out.

In the late-middle Cambrian many kinds of life-forms exterminated [59].

## 3.3. Collisional origin of modern atmosphere

1). The level of oxygen didn't change in the Phanerobiotic

According to Berner and Canfield [26], Falkowski [27], in the early Cambrian period $0_2$ concentration suddenly increased, and reached the current level already. The result indicated that life is not the main reason of atmospheric change. This mode is hard to be explained by any other hypothesis.

2). Collisional increase of ozone layer in Early Cambrian

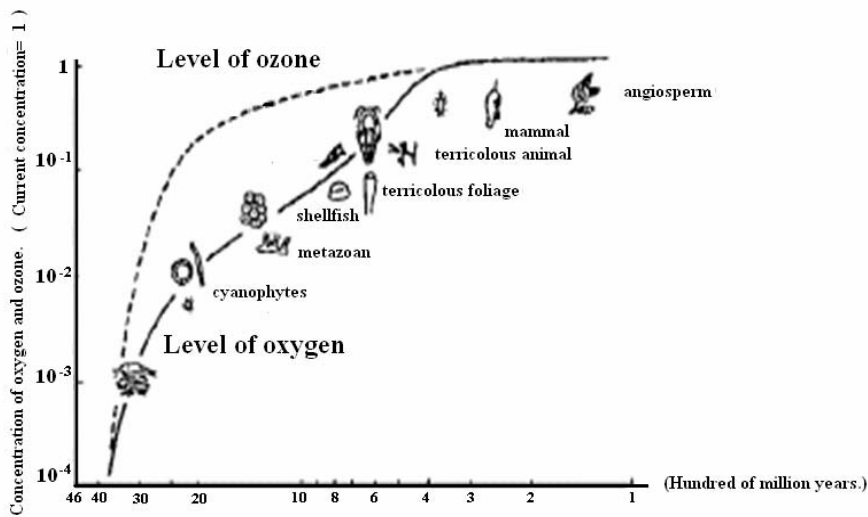

Figure 4: The curve of oxygen and ozone's level in Earth history from Iwasaka [28]. The point where the slope rapidly changed is at about 500~600Ma. The ozone layer can be formed by thermal decomposition in a high temperature. In a high temperature, many free oxygen would generated with O, there must be $O_2+O=O_3$, that's the origin of the ozone layer.

3). Carbonate rock's thermal decomposition

According to Berner [60], in the Early Cambrian, the concentration of $CO_2$ in the atmosphere is about twice of the current level, but in the Late Cambrian, the ratio increased to 18. This may be the direct effect of impact.

The impact caused the carbonate rock's thermal decomposition and generated $CO_2$.

## 3.4. Collisional parametric change of ancient Earth

1). The axis of rotation changed in the Cambrian period——True Polar Wander

From early Cambrian to middle Cambrian, the main continents on the Earth circumrotated about 90 degrees in 15 million years. The magnetic pole and the rotational pole departed, pushing the continents to the ancient equator. The axis of rotation changed [61、62、63].

If the axis didn't change, the rotation mode of the Earth would be wiggly. Since the giant impact caused the excursion of centroid, the centroid was not on the axis of rotation, the Earth wouldn't be steady.

The rotation like the right ball is also anisomerous. According to parallel axis theorem, $J = J' + md^2$, the rotational energy would be much higher.

2). The equator changed in the Cambrian period

Boucot [64] plotted three climate maps based on ancient rocks and coal mines, each represents Early Cambrian, Cambrian and Early Ordovician period. He found the tropical was moving north, indicating the change of ancient equator, echoes to the true polar wonder.

3). The temperature of ancient Earth's surface was much higher

The heat radiation of the ancient Sun is weaker than the current value.

The suggested temperature calculated on the radiation of the Sun：

Table.1. Radiation of the Sun in different period. (Cited from Salop, 1977)[65]

| Time in Earth's history | 4500Ma | 3000Ma | 1500Ma |
| --- | --- | --- | --- |
| The radiation of Sun (current value=1) | 0.85 | 0.92 | 1.00 |
| Temperature of the Earth's surface | -17 ℃ | 5 ℃ | 15 ℃ |

But the measured values were much higher: the average temperature of the Earth's surface is 70℃ in the Archaeozoic era, 52℃ in the late Proterozoic era, and 20℃ in the late Paleozoic era [66]. The result indicated that the ancient Earth is closer to the Sun. Veizer [67] also noticed that it is hard to explain the climate change in Earth history only by the change of $CO_2$'s concentration.

4). Slowdown of the Earth's rotation can be caused by an impact

In the early Cambrian, the rate of the rotation decreased rapidly. Contrary to the prediction from tidal research, the variation curve of the hours contained by each day is far above the average curve since 1300Ma. The measured slope is more inclining. The rapid slowdown in the since 1000Ma had prolonged 8.2 hours per day [68].

5). Trace on the Moon

During the Cambrian period, the Moon suffered much more meteors [69].

The velocity of Moon's leaving should increase in Cambrian period [69], because of the meteors generated in the impact would impact the Moon.

6). Collisional climate change

The Neoproterozoic global glaciation layers always accreted with the carbonate cap which represented a warm climate [20]. The carbonate cap represented the deposition of shallow sea in a warm climate. They tightly accreted, indicating a sudden climate change.

Furthermore, the thickness of the carbonate cap reached 400m, while the time span is only thousands of years.

Hence it is a rapid deposition in a high temperature.

# 4 Chemical and thermochemical Calculations

The principle of proving the atmospheric process is clear. If the modern atmosphere generated from the thermal decomposition in the impact, the temperature and pressure obtained from rock's thermal decomposition should be the same with the ones obtained from oxygen's sudden increase and the genesis of the ozone layer.

Sleep [21、22] has modeled the impact of planetesimals--objects as large as 500 kilometers across which were very common in the early Solar System. He found that a huge amount of rock would have been vaporized, lifting the surface temperatures to 3,000° C, turning the oceans into gas and driving off any atmosphere. Such conditions would have been catastrophic for any living things on the land surfaces or in the water. Sleep estimated that the Earth would have taken 2,000 years to restabilise after each impact.

## 4.1. The temperature and pressure obtained from rock's thermal decomposition

**Data 1):** In Early Cambrian, the level of $CO_2$ is about twice of current value, but in Late Cambrian, the ratio increased to 18 [60]. The paper suggested the weight of Cambrian atmosphere is close to current data (~$5.3 \times 10^{18}$kg), then the $CO_2$ created was about $2.703 \times 10^{16}$ kg. ($5.3 \times 10^{18}$(0.51%-0.06%) =$2.703 \times 10^{16}$ kg)

An estimate of the vaporized magma's amount: $n \approx n_{CaCO_3} \times 20 = n_{CO_2} \times 20 = 1.23 \times 10^{19} mol$

According to Sleep [21, 22]'s result, the author suggested the impact directly influenced an area with a radius of 600km. In such an area, the atmosphere would participate in the chemical reaction, and the seawater would vaporize.

**Reaction and formula 1):** $CaCO_3 = CaO + CO_2$

$-RTlnK \approx \Delta fHm - T\Delta fSm$

Table.2. Thermochemical data from "Thermochemical Data of Pure Substances" [70].

| Substance | $\triangle_f H_m$(J·$K^{-1}$mol) | $\triangle_f G_m$(J·$K^{-1}$mol) | $S_m$ (J·$K^{-1}$mol) |
|---|---|---|---|
| $CaCO_3$ | -1206.9 | -1128.8 | 92.9 |
| CaO | -635.1 | -604.0 | 39.75 |
| $CO_2$ | -393.5 | -394.36 | 213.64 |

**Result 1):** Above 3000℃ $CaCO_3$ will decompose completely ($K > 10^6$), thus $T \geq 3907K$

In the Archeozoic atmosphere, the concentration of $CO_2$ and $N_2$ were both 50%. In the Precambrian, there was almost no $CO_2$, while the level of $N_2$ was close to 100% [71]. According to Data 1, the total amount of atmosphere in the effected area should be $4.19 \times 10^{17}$mol.

The volume wouldn't change significantly because of the gravity, so the new amount of the gas in the area was 424 times of the former ($1.23 \times 10^{19}$mol from the gasified rock, $4.19 \times 10^{17}$mol from the pristine atmosphere of the effected area, $1.698 \times 10^{20}$mol from the gasified water). The total amount was $1.775 \times 10^{20}$mol.

According to $P = \frac{nRT}{v}$, the air pressure of the effected area was about 5600Bar.

(The whole process can be viewed in my website: http://218.108.56.172/others/zwj/down/10.doc)

## 4. 2. The excellent accordance of temperature data and the origin of the ozone layer

**Data 2):** The Precambrian atmosphere has little oxygen according to Kimura and Watanabe [59]. The author synthesized results from Berner and Canfield [26] and other researchers. In the middle Cambrian the concentration of oxygen suddenly increased and reached 3%-10% of current value. The author chose the average number 6.5%. First we don't need to care in which form would the oxygen atoms exist, $m_{oxygen-atoms} = 7.93 \times 10^{16} kg$, $n_{oxygen-atoms} = 4.96 \times 10^{18} mol$.

The well-accepted view told us that the early Cambrian Earth was covered by a large shallow ocean. In a $600^2 \pi$ sq.Km. area, there should be $3.06 \times 10^{18} kg$ water. When the water boiled away, there should be $1.698 \times 10^{20} mol$ vapor ($H_2O$) in the atmosphere.

Table.3. Thermochemical data in this reaction from "Thermochemical Data of Pure Substances" [70].

| Substance | $\triangle_f H_m$(J·$K^{-1}$mol) | $\triangle_f G_m$(J·$K^{-1}$mol) | $S_m$ (J·$K^{-1}$mol) | $C_{m,p}$ (J·$K^{-1}$$mol^{-2}$) |
|---|---|---|---|---|
| $H_2O$ | -241.82 | -228.59 | 188.72 | 33.58 |
| $O_2$ | 0 | 0 | 205.03 | 29.36 |
| O | 247.52 | 230.09 | 160.95 | 21.91 |
| $H_2$ | 0 | 0 | 130.59 | 28.84 |

**Reaction and formula 2):** $2H_2O = 2H_2 + O_2$ (Normal decomposition), $H_2O = H_2 + O$ (Free radical decomposition in a high temperature), $O_2 + O = O_3$



$$\Delta rHmT = \Delta rHm^\Theta + \int_{298.15}^{T} \Delta CpdT, \quad \Delta rSmT = \Delta rSm^\Theta + \int_{298.15}^{T} \frac{\Delta Cp}{T} dT$$

**Result 2)**: The critical temperature of normal decomposition is 5450K, while the one of free radical decomposition is 4759Km (much lower). So the decomposition of water would also produce free oxygen atoms at 4000K! Supposing all the water reacted as normal decomposition, T=3432K; then if all the water reacted as the free radical decomposition, T=4111K. The temperature obtained from different methods matched well. The temperature was about 4000K, while the pressure was about 5600Bar.

（The whole process can be viewed in my website： http://218.108.56.172/others/zwj/down/10.doc）

Since the average temperature was so high, and many free oxygen had generated, there must be the chemical reaction $O_2+O=O_3$. That's just the origin of the ozone layer.

**Data 3)**: In the standard condition, for $O_3$, $\Delta rH_m = 142200J$, $\Delta rS_m = 237.6J$, $C_{P,m} = 38.16J$. Convert to 3772K, then use $-RTlnK = \Delta H - T\Delta S$ and got K=$4.418\times 10^{-7}$,

The simplified expression of K $\Rightarrow nO_3 = 1.401\times 10^{-23} \times nO \times 2nO_2$. The sum of nO and $2nO_2$ is a constant, nO + $2nO_2$=$4.96\times 10^{18}$ mol, so $nO_3 \leq 1.401\times 10^{-23} \times (2.48\times 10^{18})^2 = 8.61\times 10^{13}\, mol$

**Result 3)**: While the current measured value of $nO_3$ is $6.8\times 10^{13}$mol, just fits our result. (If compress to 1atm at 273.15K, the thickness will be 0.3cm). The thermal decomposition in the impact may be the origin of ozone layer.

# 5  The cyber-simulation by EQS4WIN

The simulation will adopt EQS4WIN, programmed by Mathtrek group. The author put hundreds of possible atmospheric components and all possible chemical changes into the program. Then the author put into the temperature, air pressure and other parameters of ancient atmosphere got from the calculation above, and compared the result with the predicted value.

Data chosen were measured by experiments in high temperature from "Thermochemical Data of Pure Substances" [70], and were converted to data under 4000K using formula

$$\Delta rHmT = \Delta rHm^\Theta + \int_{298.15}^{T} \Delta CpdT, \quad \Delta rSmT = \Delta rSm^\Theta + \int_{298.15}^{T} \frac{\Delta Cp}{T} dT.$$

Table.4. Part of the thermochemical data of atmospheric components from "Thermochemical Data of Pure Substances" [70].

| Gas | T | Cp | S | -(G-H298)/T | H | H-H298 | G | ΔHf | ΔGf | logKf |
|---|---|---|---|---|---|---|---|---|---|---|
| | K | J/mol.K | | | KJ/mol | | | | | |
| HNO3（g） | 2200 | 103.858 | 431.700 | 353.608 | 37.498 | 171.804 | -912.243 | -124.275 | 316.752 | -7.521 |
| H2O2（g） | 1500 | 68.328 | 324.479 | 276.547 | 64.207 | 71.899 | -550.926 | -141.095 | 27.556 | -0.960 |
| HNCO（g） | 3000 | 80.148 | 388.922 | 324.469 | 91.687 | 193.358 | -1075.079 | -108.350 | 8.697 | -0.151 |
| CH2 | 3000 | 57.182 | 298.575 | 253.610 | 521.285 | 134.893 | -374.439 | 372.243 | 237.633 | -4.138 |
| CH4（g） | 2000 | 94.420 | 305.811 | 44.614 | 48.721 | 123.594 | -562.901 | -92.504 | 130.940 | -3.420 |
| NH（g） | 3000 | 38.006 | 255.185 | 224.711 | 467.980 | 91.420 | -297.574 | 377.252 | 316.373 | -5.509 |
| NH2（g） | 3000 | 58.186 | 294.543 | 251.231 | 320.311 | 129.939 | -563.320 | 185.213 | 310.595 | -5.408 |
| NH3（g） | 3000 | 78.938 | 322.414 | 264.104 | 128.990 | 174.931 | -838.251 | -50.477 | 295.630 | -5.147 |
| C2O（g） | 3000 | 66.520 | 362.742 | 308.586 | 453.511 | 162.468 | -634.715 | 283.899 | -72.747 | 1.267 |
| N2H4（g） | 1600 | 111.109 | 376.890 | 304.768 | 210.581 | 115.395 | -392.443 | 89.594 | 456.214 | -14.894 |
| OH（g） | 4000 | 37.885 | 267.561 | 235.827 | 165.926 | 126.939 | -904.320 | 33.136 | -17.091 | 0.223 |

| | | | | | | | | | |
|---|---|---|---|---|---|---|---|---|---|
| H2O (g) | 4000 | 58.033 | 303.009 | 257.121 | -58.275 | 183.551 | -1270.309 | -254.501 | -18.821 | 0.246 |
| CH2O (g) | 3000 | 79.575 | 355.258 | 294.189 | 67.307 | 183.208 | -998.467 | -130.742 | -8.703 | 0.152 |
| CH | 3000 | 41.382 | 260.645 | 227.920 | 692.302 | 98.174 | -89.633 | 587.629 | 262.472 | -4.570 |
| CaO (g) | 3500 | 63.514 | 323.429 | 278.956 | 199.587 | 155.655 | -932.415 | -106.647 | -5.915 | 0.088 |
| CaCO3 | 1200 | 130.541 | 244.634 | 159.075 | -1104.25 | 102.671 | -1397.811 | -1203.73 | -901.388 | 39.236 |
| NO (g) | 3000 | 37.469 | 288.170 | 256.512 | 185.266 | 94.975 | -679.245 | 89.902 | 52.427 | -0.913 |
| NO2 (g) | 3000 | 57.584 | 354.995 | 306.912 | 177.343 | 144.248 | -887.642 | 32.972 | 221.723 | -3.861 |
| N2O (g) | 3000 | 61.693 | 341.441 | 290.480 | 234.928 | 152.880 | -789.393 | 93.207 | 296.259 | -5.158 |
| N2O3 (g) | 3000 | 103.028 | 515.557 | 429.415 | 341.269 | 258.426 | -1205.401 | 101.534 | 635.635 | -11.067 |
| N2O4 (g) | 3000 | 131.781 | 564.457 | 454.865 | 337.856 | 328.717 | -1355.515 | 49.114 | 863.214 | -15.030 |
| N2O5 (g) | 3000 | 148.705 | 655.749 | 528.036 | 394.436 | 383.139 | -1572.810 | 56.688 | 1023.610 | -17.823 |
| NO3 (g) | 3000 | 82.414 | 418.193 | 348.716 | 279.558 | 208.430 | -975.019 | 86.180 | 512.037 | -8.915 |
| CN (g) 398 | 3000 | 42.010 | 280.343 | 247.930 | 532.373 | 97.237 | -308.655 | 425.713 | 137.462 | -2.393 |
| CN2 (g) 399 | 3000 | 61.807 | 353.329 | 300.779 | 630.441 | 157.649 | -429.596 | 477.423 | 370.551 | -6.452 |

Other data were computed from the data at 298K.

According to $\Delta G = \Delta H - T\Delta S$, the author input the Gibbs free energies at 4000K, the temperature, the air pressure and the components of the atmosphere before the impact.

The result of the simulation indicated the genesis of pure $O_2$ and the escape of $H_2$. Here are the components of the atmosphere (influenced area) after the impact:

$O_2$: $5.83 \times 10^{18}$ mol; O: $1.26 \times 10^{18}$ mol; $NH_3$: $1.97 \times 10^{14}$ mol; $O_3$: $7.24 \times 10^{13}$ mol;

The result of the simulation also indicated a preservation of pure oxygen. The hydrogen would escape at 4000K. The diffusion velocity of hydrogen is extremely fast. At 1300K, the average velocity of hydrogen will reach 5000m/s. Furthermore, the cyber-simulation indicated that lots of the hydrogen would decompose and become free radicals. But at such an extreme condition, the Gibbs free energy of the reaction $N_2 + 6H = 2 NH_3$ reached -1252700J/mol, which meant the reaction will continue until the last molecular! (The $O_2$ couldn't oxidate $N_2$ at the same condition.)

# 6 Conclusions

Large amount of evidences supported such a hypothesis:

In Late Precambrian, a celestial body impacted the Earth. The impact solved many problems and was supported by a large amount of evidences. The high temperature ended the great glaciation, facilitated the communication of biological information. The rapid changes of Earth environment enkindled the genesis-control system, and released the HSP-90 variations. After the impact, benefited from the protection of the new ozone layer and the energy supplement of the aerobic respiration, those survived underground life exploded. They generated carapaces and complex metabolism to adjust to the new circumstance of high temperature and high pressure. Maybe one of those swing arms has an abnormal gravitational origin. That's the reason of the impact. The hypothesis may be the solution to the puzzles of the Early Cambrian.

The author did chemical calculation on the variation of atmospheric components.

And the hypothesis is in accordance with the result of computer simulation.

# Acknowledgements

I shall gratefully acknowledge my dear teacher Prof. LIU Xiao-han (Chinese Academy of Sciences) for his help and comments. He helped me a lot and I am very grateful.

Very thanks to my good friend, dear teacher Prof. CHEN Yan-jing (School of earth and space sciences, Peking University) for encouragement and help. Discussions with him were particularly helpful.

I acknowledge Prof. PAN Mao (Dean of the School of earth and space sciences, Peking University), Prof. NING Jie-Yuan (School of earth and space sciences, Peking University) and Prof. WANG Chang-Ping (Dean of the school of mathematical sciences, Peking University) for their valuable comments on an earlier manuscript. Comments from Prof. S. G. HAO (School of earth and space sciences, Peking University) helped improve the manuscript and are gratefully acknowledged. The author also wishes to thank Mr. JIANG Wang-qi (Peking University) for his help.

The research was granted by National Undergraduate Innovative Test Program.

# Author Brief Introduction

Weijia Zhang was born in Hangzhou, China in December 4, 1989. He studied physics, chemistry, paleontology and paleogeology in Peking University, and stays there until now. He received the title of "Future Scientists" which was granted by the Department of Education and the Chinese Academy of Sciences at the age of 16. Only three young scientists in China were granted.